\documentclass[%
reprint, superscriptaddress,amsmath,amssymb,aps,
prl,prstper,floatfix,showpacs]{revtex4-1}

\usepackage[FIGTOPCAP,tight]{subfigure}
\usepackage{amssymb}
\usepackage{amsmath}
\usepackage{graphicx}
\usepackage{array}
\usepackage{epstopdf}
\usepackage[usenames, dvipsnames]{color}

\newcommand{\pare}[1]{\left( #1 \right)}

\newcommand{\cor}[1]{\left[ #1 \right]}

\newcommand{\ave}[1]{\left\langle #1 \right\rangle}

\begin{document}

\title{Phase dependent vectorial current control in symmetric noisy optical ratchets}

\author{Magda G. S\'anchez-S\'anchez}
\address{Instituto de Ciencias Nucleares, Universidad Nacional Aut\'{o}noma de M\'{e}xico, Apartado Postal 70-543, 04510 Cd. Mx., M\'{e}xico}

\author{Roberto de J. Le\'{o}n-Montiel}
\email{roberto.leon@nucleares.unam.mx}
\address{Instituto de Ciencias Nucleares, Universidad Nacional Aut\'{o}noma de M\'{e}xico, Apartado Postal 70-543, 04510 Cd. Mx., M\'{e}xico}

\author{Pedro A. Quinto-Su}
\email{pedro.quinto@nucleares.unam.mx}
\address{Instituto de Ciencias Nucleares, Universidad Nacional Aut\'{o}noma de M\'{e}xico, Apartado Postal 70-543, 04510 Cd. Mx., M\'{e}xico}


% 05.40.Ca  Noise, fluctuation phenomena
% 05.40.-a  Fluctuation phenomena, statistical physics
% 05.60.Cd  Transport processes, classical
% 82.70.Dd  Colloids
% 87.80.Cc  Optical cooling and trapping in biophysics
% 87.80.Cc	Optical trapping (see also 42.50.Wk Mechanical effects of light on material media, microstructure and particles in optics; 37.10.-x Atom, molecule, and ion cooling methods)

\begin{abstract}
In this work we demonstrate single microparticle transport in a symmetric noisy optical ratchet where each potential is a low power ($<2.5$ mW) three dimensional trap. The optical potentials consist of 20 symmetric optical traps arranged in a one-dimensional lattice produced by a spatial light modulator. The external periodic force of the ratchet system adds to zero over one period (symmetric) and is generated by the motion of a piezo-electric microscope stage. Transport is achieved by adding noise to the potentials by randomly varying the diffracted power into the traps at the same frequency of the external force. We show that the direction and speed of motion (current) is coupled to the phase difference between the noise in the optical potentials and the external periodic force.
\end{abstract}

\maketitle

Transport phenomena are ubiquitous to a wide variety of disciplines ranging from biology, chemistry, physics, and even electronics \cite{rebentrost2009,caruso2009,roberto2013,spiechowicz2014,roberto2014,spiechowicz2015,leon2015,spiechowicz2016,biggerstaff2015}. In particular, directed motion of particles has become a major research topic \cite{hanggi2009}, mainly because of its fundamental role in understanding the functioning of natural systems, such as the movement of motor proteins along tubulin molecules \cite{ajdari1991,simon1992,leibler1994}, which has lead to novel applications ranging from microparticle sorting \cite{huang2004}, to the directed motion of cold \cite{gomers2006,gomers2008} and ultracold atoms \cite{salger2009}, to the control of electronic transport through semiconductor superlattices \cite{bass_book}.

Interestingly, unidirectional motion in the micro- and nanoscale has been achieved by means of ratchet systems, where the movement of a particle is mediated by a fine-tuned combination of a zero-average periodic external force and asymmetric potentials which privilege motion in one direction while hindering it in the opposite \cite{faucheux1995,lee2005PRL,arzola2011,hasegawa2012,huidobro2013,arzola2017}. These asymmetric potentials represent the ratchet and the pawl in the classical Smoluchowski-Feynman ratchet \cite{smoluchowski1912,feynman1963,hanggi1996}, while the periodic force represents the Brownian perturbations. In striking contrast, in the absence of spatial asymmetry, conventional wisdom says that a zero-average periodic external destroys directed transport \cite{reimann2001}, and new approaches need to be considered \cite{flach2000,zheng2001}. Among these, the simplest strategy consists of introducing a non-zero, constant, strong tilting force to the potentials, which drives the system out of equilibrium, thus producing movement of the particle \cite{reimann2002}. Another approach is to use a weak tilting force---much smaller than the one necessary to make the particle escape the potentials---in combination with dynamically-disordered potentials \cite{leon2017}. Finally, the third and most difficult way of producing directed motion is by incorporating correlated noise, rather than white or Brownian, to the dynamics of the particle \cite{luczka1995}.

In this work, we show theoretically and experimentally that directed motion of a single particle can be observed in systems where both the potential and the zero-average driving force are symmetric. This is obtained by introducing simple Gaussian white noise to the potentials, whose phase, relative to the periodic force, determines the direction and the speed of the particle's movement. Because of its simplicity, and versatility, our experiment constitutes a robust platform for the study of directed motion across symmetric potentials, and paves the way towards the development of novel noise-enabled micro- and nanoscale transport technologies.

\begin{figure}[t!]
   \begin{center}
   \includegraphics[width=8.5cm]{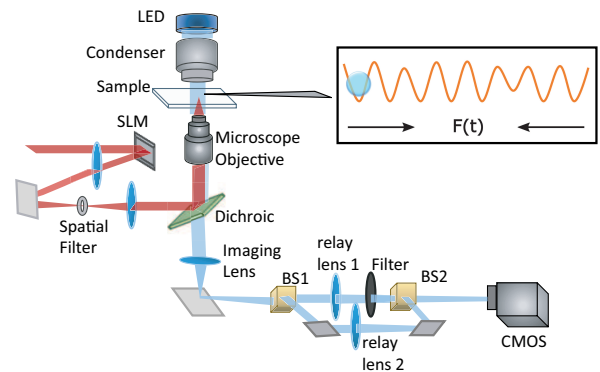}  
   \end{center}
   \vspace{-3mm}
   \caption{Experimental setup: The expanded trapping laser beam is reflected at an SLM that imprints the 2d phase to create a linear array of 20 traps. The laser beam is focused by the microscope objective in the liquid sample which is illuminated by an LED. The imaging system is divided by the first beam splitter (BS1) into two paths: one that blocks the trapping light with a filter and projects (relay lens 1) the image of the trapped particle and another that projects (relay lens 2) the reflected trapping laser light as a speckle pattern. Both paths are recombined by the second beam splitter (BS2) at the camera. Inset: Schematic representation of the symmetric noise-enabled optical ratchet. }
\end{figure}

The experimental setup is depicted in Fig. 1. A single microparticle is trapped by one of the optical potentials contained in a linear array of traps. The microparticle sample is composed of silica microbeads (Bangs Laboratories, mean diameter $2R$ of 2.47$~\mu$m) immersed in water and contained within two microscope coverslips (No. 1, 0.13-0.16 mm thick) separated by $\sim 100~\mu$m. The symmetric ratchet is formed by a linear array of dynamical optical potentials and a symmetric external force induced by the movement of the microscope stage. The linear array of 20 optical traps is created by simultaneously focusing the trapping laser beam (1064 nm), which is shaped by a spatial light modulator (SLM, Holoeye Pluto NIR 2) before entering the microscope objective (100x/1.25NA). The resulting linear array of focused spots are separated by a distance of $d_t=2.3\pm 0.1~\mu$m.
%%%%%%%%%%%%%%%%%%%%%%%%%%%%%%%%%%%%%%%%

The events in the experiment are imaged by a CMOS camera (Thorlabs, DCC1545M) that records at 40 frames per second (fps). The imaging system transmits the LED illumination and the reflected trapping laser, both components are directed through two paths (Fig. 1): The first path images the trapping plane through the first relay lens, where the laser light is blocked by a filter (low pass, cutoff at 950 nm).
The second path has no filter so that the light from the reflected laser is projected (relay lens 2) as a speckle pattern. Both paths are recombined at the CMOS sensor. Note that this configuration allows us to measure, at the same time, the position of the trapped particle, a reference particle at the bottom to track the movement of the microscope stage, and the change in the trapping pattern when the digital hologram at the SLM is updated.

%%%%%%%%%%%%%%%%%%%%%%%%%%%%%%%%%%%%%%%%%%%%%%%%%%%%
In our experiments, both the external periodic force and the changes in the optical potentials are driven at a frequency $f$. We set the frequency considering the following: (1) A particle trapped in one of the potentials has to move a distance $d$ to be pulled by a neighboring potential. If the depth of the potentials is the same, then $d_t/2<d<d_t$ (particle has to move more than half the distance to the next potential). (2) In the case of a tilted ratchet (constant  external force) where $d\sim R$ \cite{leon2017}, transport is enhanced when the noise correlation time $1/f$ is similar to the particle configurational relaxation time $\tau _{cr}\sim \eta R^3 /(2 k_B T)$, where $\eta$ is the liquid viscosity. $\tau _{cr}$ is the time needed for the particle to diffuse across its own radius. Here $d> R$, so the  time required for the particle to diffuse across $d$ is $\tau _{cr2}\sim \eta R d^2/(2 k_B T)$. In our case the condition $d_t/2<d<d_t$ yields $\tau _{cr2}$ in the range between 0.36 and 0.7 s, resulting in frequencies between 1.4 and 2.8 Hz. Hence we set the frequency of the external force and noise at the mean value of $f=2$ Hz.

The external periodic force is introduced by moving the piezo-electric  microscope stage (Thorlabs, Max301) which is driven by a computer-controlled function generator that outputs a symmetric triangular wave so that the speed of the stage is constant $23.8\pm 1.6~\mu$m$/$s. In this way, the trapped bead experiences the drag force induced by the motion of the piezo-electric stage. This force can be calculated by means of the Stokes expression $F_d = 6 \pi \eta R v$, where $\eta$ is the liquid viscosity and $v$ is the drag speed. Because the microparticle is trapped at a height $h$ of $12 \mu$m above the bottom coverslip, the Stokes formula has to be corrected with the Faxen correction in ratios of $r=R/h$ \cite{leach2009}. In this case the contribution of the correction is  $6\%$ compared to the Stokes expression. It is important to highlight that the magnitude of the constant external force of $23.8\pm 1.6~\mu$m/s is not enough to induce transport with a static pattern of optical potentials, because the external force is much smaller than the escape velocity of the optical potentials ($49.4 \pm 17.8 ~\mu$m).

%%%%%%%%%%%%%%%%%%%%%%%%%%%%%%%%%%%%%%%%%%%%%%%%%%%%%%%%%
\begin{figure}[t!]
   \begin{center}
   \includegraphics[width=8.5cm]{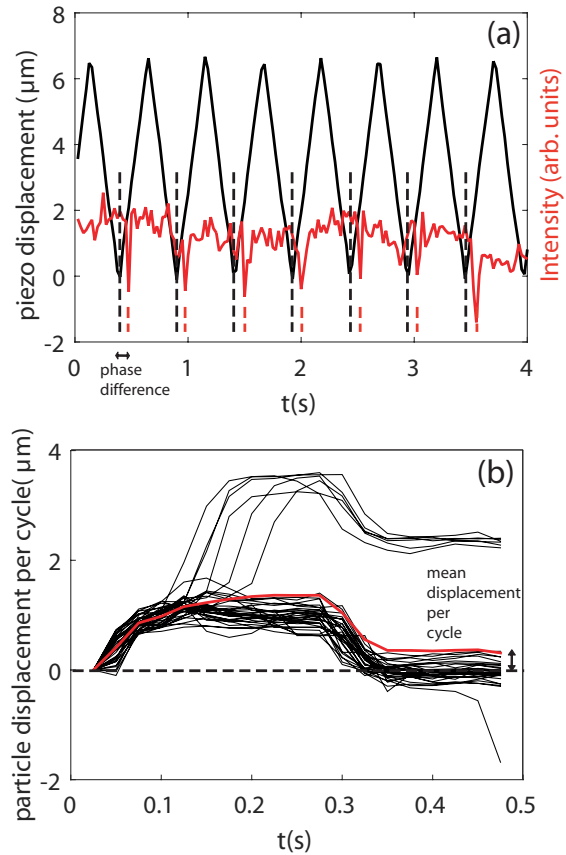}   
   \end{center}
   \vspace{-3mm}
   \caption{(a) Top: Measured trajectory of a reference microparticle stuck at the bottom of the microscope coverslip. The red line shows the change in the hologram. The phase difference is $1.1$ rad. (b) Bottom: Measured trajectories of the trapped microparticle for each cycle. The red line represents the average trajectory over 37 different realizations for a phase difference of $1.1$ rad. }
\label{Fig:eff}
\end{figure}
%%%%%%%%%%%%%%%%%%%%%%%%%%%%%%%%%%%%%%%%%%%%%%%%%%%%%%%%%%%%

Microparticle transport is achieved by introducing dynamical disorder or noise into the optical potentials by means of random changes in the optical power at the individual traps. This is done by changing the projected digital holograms at the same rate as the external force (2 Hz). The variation  in the diffracted power to the individual traps has a standard deviation of $\pm 23\%$. The digital holograms are calculated with the simple Gershberg-Saxton algorigthm \cite{gs}, randomly changing the target intensity at the potentials. The total diffracted power into all of the potentials is 37 mW (transmitted by the microscope objective), which results in a measured escape velocity for different projected patterns of $49.4\pm 17.8~\mu$m/s. Here, it is worth mentioning that each time the hologram is changed there is flickering, which results in a lowering of a diffracted power to the trap. In our experiment, we found that during a hologram-switching event the particle is essentially free for about 5 ms; this results in a displacement of 119 nm, which is not enough to reach a neighboring potential well.

%%%%%%%%%%%%%%%%%%%%%%%%%%%%%%%%%%%%%%%%%%%5

\begin{figure}[t!]
   \begin{center}
   \includegraphics[width=8.5cm]{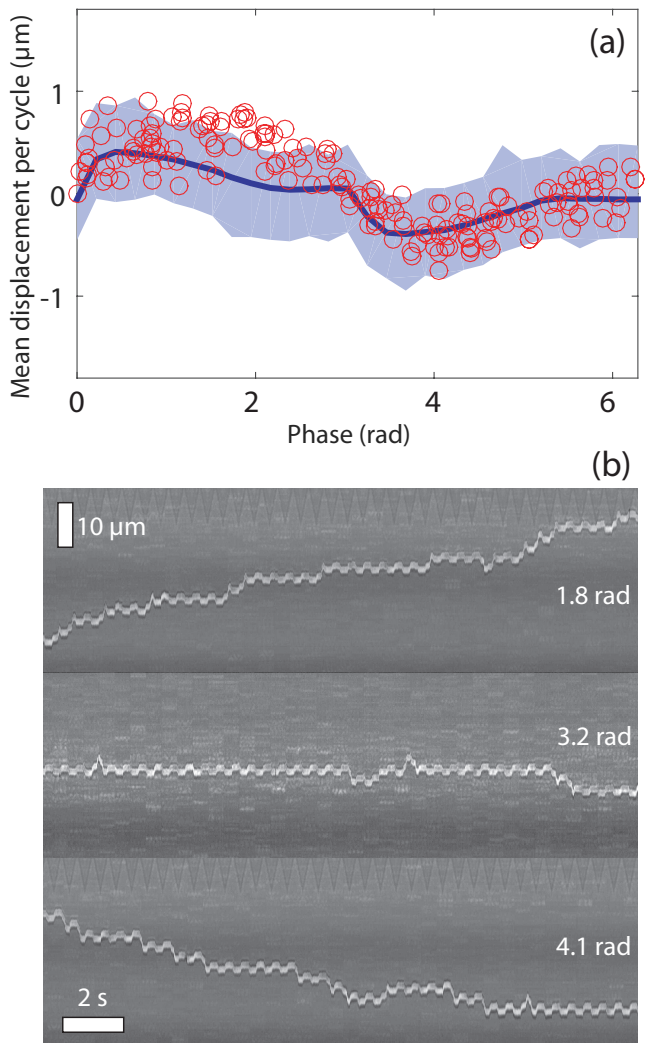}
   \end{center}
   \vspace{-3mm}
   \caption{(a) Average displacement (per cycle) of the particle as a function of the periodic-force phase $\phi$. The red round markers correspond to 37 different realizations of the experiment. The blue shaded region is composed by 1000 numerical realizations, showing that the measured signals can be described by our theoretical model. (b) Composite images for phase differences of 1.8, 3.2 and 4.1 rad for a time of 19.5 s. Each video frame is cut to a width of 3 pixels and each composite image contains 780 frames. }
\label{Fig:displacement}
\end{figure}

%%%%%%%%%%%%%%%%%%%%%%%%%%%%%%%%%%%%%%%%%%%%%%%%%%%%%%%%%%%%
Figure 2(a) shows the displacement of the piezo stage (black triangular line-reference microparticle) and the intensity of the speckle pattern (red line), which depicts the period of time at which the digital hologram changes. In the case of the external force, we can see that the slope (proportional to the force) depends on the amplitude of the stage motion. In this case the amplitude is $6.1 \pm 0.3 ~\mu$m and a speed of $23.8\pm 1.6 ~\mu$m/s. We define the zero of the phase at the lowest amplitude of the piezo-stage signal.
%%%%%%%%%%%%%%%%%%%%%%%%%%%%%%%%%%%%%%%%%%%%%%%%%%%%%%%%%%%%%%%
Figure 2(b) shows the displacement of the trapped particle for a single phase difference of $1.1$ rad in 37 cycles (black lines), and the average displacement (red line). Notice that after one cycle the particle does not return to its initial position but a finite, small distance above it. This phase-controlled movement within one cycle is what we define as the mean displacement per cycle.
%%%%%%%%%%%%%%%%%%%%%%%%%%%%%%%%%%%%%%%%

%%%%%%%%%%%%%%%%%%%%%%%%%%%%%%%%%%%%%%%%%%%%%%%%%
Each experiment is performed in the following way: a particle is trapped by one of the potentials and raised to a height of $12~\mu$m. Then the video of digital holograms is projected (2Hz) into the SLM and we confirm that in the absence of external force there is no transport. The next step is to pause the SLM and introduce the external force (2Hz) of the microscope stage to check that there is no transport with static potentials either. The transport experiment is then started with both the SLM video and the periodic external force. The phase difference between both signals is controlled by varying the phase of the waveform generated by the function generator (start at zero). The phase in the function generator is changed in steps of 40 degrees ($\sim 0.70$ rad). In order to avoid the particle reaching one of the boundaries (initial and last traps), each phase is switched every 20 seconds by $\pi$ radians (the phase difference at which the direction of movement is turned over). In this way we start with zero radians followed by $\pi$, 0.7 , 3.8..., at the function generator.

%%%%%%%%%%%%%%%%%%%%%%%%%%%%%%%%%%%%%%%%%%%%%%%%%%%
%%%%%%%%%%%%%%%%%%%%%%%%%%%%%%%%%%%%%%%%%%%%%%%%%%%%

The results of our experiments are depicted in Fig. 3. In Fig. 3(a)  we plot (circles) the average displacement per cycle as a function of phase difference. Each data point represents the average over 37 cycles. The shaded area is the result of the theoretical description of our experiment, which we describe below. Figure 3(b) shows composite images extracted from the video data at three different phases. To make the image, we take a slice with a width of 3 pixels for each video frame. The particle center appears bright, so that the trajectory for more than 19 seconds looks like a white line in the composite image.

%%%%%%%%%%%%%%%%%%%%%%%%%%%%%%%%%%%%%
To understand the results of our experiment, we make use of a minimalistic model in which the silica microbeads are assumed to be Brownian particles moving in a dynamically-disordered (noisy) one-dimensional potential lattice, as depicted in the inset of Fig. 1. The system is composed by $N$ closely-spaced Gaussian wells described by \cite{lee2005,pelton2004}
\begin{equation}
V\pare{x}=-V_{0}\pare{t}\sum_{n=0}^{N}\exp\cor{-\frac{\pare{x-nL}^{2}}{2\sigma^2}},
\label{Eq:potential}
\end{equation}
with $V_{0}\pare{t}$ and $\sigma$ describing the depth and width of the wells, and $L$ the separation between them.
%
%
%\\
In the experiment, noise is introduced as random variations in the power at each optical trap (trap depth). This can be captured in our theoretical model by introducing stochastic fluctuations in the depth of the wells, which can be written as $V_{0}\pare{t} = V_{0}\cor{1 + \phi\pare{t}}$, where $V_{0}$ stands for the average depth of the wells and $\phi\pare{t}$ is a Gaussian random variable with zero average, i.e. $\ave{\phi\pare{t}}=0$, with $\ave{\cdots}$ denoting stochastic averaging. Note that the frequency of the noise is defined by the inverse of the time that takes our noise generator to change the value of the $\phi$ variable.

The motion of the Brownian particle due to the external periodic force $F\pare{t}$, can be well described by the Langevin equation, in the overdamped limit \cite{volpe2012}, as
\begin{equation}
\dot{x} = -\frac{1}{\gamma}\cor{V\pare{x}+F\pare{t}} + \sqrt{2k_{B}T\gamma}\xi\pare{t}.
\label{Eq:Langevin}
\end{equation}
Here, $\gamma$ characterizes the friction of the particle immersed in liquid, and $\sqrt{2k_{B}T\gamma}\xi\pare{t}$ the thermal noise due to random collisions with the surrounding fluid molecules. $\xi\pare{t}$ stands for a Gaussian Markov stochastic process with zero average, $k_{B}$ is the Boltzmann constant, and $T$ the temperature of the system. The external periodic force is given by
\begin{equation}
F\pare{t} = F_{d}\Omega\pare{2\pi\nu t + \tilde{\phi}},
\end{equation}
where $\Omega\pare{t}$ describes a zero-average periodic square-wave function, with a frequency $\nu$ and a $\tilde{\phi}$ phase shift. Here, it is important to highlight that our system constitutes a true optical ratchet, as the effective force is null, that is, the external periodic force averages to zero in the long-time limit.

By making use of the model described above, we can thus calculate the particle's average displacement per cycle in a 20-site symmetric one-dimensional lattice as a function of the phase difference between the external force and the noise signal. The shaded region of Fig. 3 shows 1000 realizations of the experimental procedure described above, with the solid line depicting the average. Note that the experimental results are well captured by our simplified theoretical model, showing that the particle's motion can indeed be controlled by properly setting the phase difference between the drag and noise signals.

In conclusion, we have presented a phase-controlled symmetric noise-enabled optical ratchet where magnitude and direction of the current can be adjusted just by changing the phase difference between noise and external force. Contrary to conventional knowledge, our results demonstrate that particle motion in symmetric potentials subjected to a periodic zero-average external force can be observed and manipulated by introducing simple Gaussian white noise to the potentials. More importantly, we show that the direction of the particle's motion can easily be controlled by changing the relative phase between the force and noise signals. This might help expanding the scope of particle transport in symmetric potentials and open the door to new noise-enabled micro- and nanoscale technologies.

\section*{Acknowledgment}
Work partially funded by the following grants: DGAPA-UNAM: PAPIIT IA100718 and IN107719, Conacyt: CB-2016-01/284372, CB253706 and LN293471. PAQS thanks Mr. Jos\'e Rangel Guti\'errez for machining custom opto-mechanical parts.

\end{document}